\documentclass[12pt,a4paper]{article}
\usepackage{a4wide,amsmath}

\newcommand{\Tr}{\mathop{\mathrm{Tr}}\nolimits}

\title{Complete analysis of polarization effects in
$\gamma\gamma\leftrightarrow e^+e^-$}
\author{V.~N.~Baier and A.~G.~Grozin\\
\textit{Budker Institute of Nuclear Physics, Novosibirsk 630090, Russia}}
\date{}

\begin{document}

\maketitle

\begin{abstract}
Formulae for the cross section of the process $\gamma\gamma\leftrightarrow e^+e^-$
with all four particles polarized are obtained in the lowest order
of the perturbation theory using REDUCE.
\end{abstract}

Polarization effects in the quantum electrodynamics process
$\gamma\gamma\leftrightarrow e^+e^-$
were discussed in many works (see e.g.~\cite{BLP,M}),
but formulae with all four particles polarized are absent in the available literature.
These polarization effects may be important for interaction of high energy
electron beams with intense laser beams and some other applications.
Here we obtain a complete set of formulae in covariant notations
using REDUCE~\cite{H}.
For photon-electron scattering (Compton effect) this was done in~\cite{G}.
We follow the method of~\cite{BLP} and~\cite{G}.

We start from the process $\gamma\gamma\to e^+e^-$.
The particles' momenta are $k+k'=p+p'$;
the Mandelstam invariants are $t=(k-p)^2=m^2(1-x)$, $u=(k-p')^2=m^2(1-y)$,
$s=m^2(x+y)$.
The cross section has the form
\begin{equation}
\begin{split}
&\frac{d\sigma}{dt\,d\varphi} = \frac{\alpha^2}{2 s^2 x^2 y^2}
\rho'_{\mu\alpha} \rho_{\nu\beta}
\Tr \rho' Q^{\mu\nu} \rho \overline{Q}^{\alpha\beta}\,,\\
&\frac{m Q^{\mu\nu}}{xy}
= \frac{\gamma^\mu (\hat{p}'-\hat{k}'+m) \gamma^\nu}{x}
+ \frac{\gamma^\nu (\hat{p}'-\hat{k}+m) \gamma^\mu}{y}\,,
\end{split}
\label{form}
\end{equation}
where $\varphi$ is the azimuthal angle in the plane
transverse to the line of collision of the initial particles.
All density matrices are defined as in~\cite{BLP}.

It is convenient to use the basis
\begin{equation}
n_0 = \frac{Q}{v}\,,\quad
n_1 = \frac{K}{v}\,,\quad
n_2 = \frac{v}{2w} P_\bot\,,\quad
n_3^\mu = - \frac{1}{2vw}
\varepsilon^\mu{}_{\alpha\beta\gamma} Q^\alpha K^\beta P^\gamma\,,
\label{basis}
\end{equation}
where $Q=k+k'=p+p'$, $K=k-k'$, $P=p'-p$, $P_\bot=P-\frac{PK}{K^2}K$;
$v=\sqrt{x+y}$, $w=\sqrt{xy-x-y}$.
The particles' momenta are
\begin{equation}
\begin{split}
&p = \frac{(x+y) n_0 - (x-y) n_1 - 2 w n_2}{2v}\,,\quad
p' = \frac{(x+y) n_0 + (x-y) n_1 + 2 w n_2}{2v}\,,\\
&k = \frac{v}{2} \left(n_0+n_1\right)\,,\quad
k' = \frac{v}{2} \left(n_0-n_1\right)\,.
\end{split}
\label{momenta}
\end{equation}
The vectors $\vec{n}_1$, $\vec{n}_2$, $\vec{n}_3$ form a right-handed system.

The vectors $n_2$ and $n_3$ can be used as polarization vectors of both photons.
For the photon with momentum $k$,
the vectors $\vec{n}_2$, $\vec{n}_3$, $\vec{k}$ form a right-handed system
(in the c.~m.\ frame);
therefore, its density matrix is expressed via the Stokes parameters $\xi_j$
in the standard way:
$\rho_{\nu\beta} = \frac{1}{2} \sum_{j=0}^{3} \xi_j \sigma_{j\nu\beta}$,
where $\xi_0=1$ has been formally introduced, and
\begin{equation}
\begin{split}
&\sigma_0^{\mu\nu} = n_2^\mu n_2^\nu + n_3^\mu n_3^\nu\,,\quad
\sigma_1^{\mu\nu} = n_2^\mu n_3^\nu + n_3^\mu n_2^\nu\,,\\
&\sigma_2^{\mu\nu} = - i \left( n_2^\mu n_3^\nu - n_3^\mu n_2^\nu \right)\,,\quad
\sigma_3^{\mu\nu} = n_2^\mu n_2^\nu - n_3^\mu n_3^\nu\,.
\end{split}
\label{sigma}
\end{equation}
For the photon with momentum $k'$,
the right-handed system is $\vec{n}_2$, $-\vec{n}_3$, $\vec{k}\,'$.
Therefore,
$\rho'_{\mu\alpha} = \frac{1}{2} \sum_{j'=0}^{3}
\delta_{j'} \xi'_{j'} \sigma_{j'\mu\alpha}$,
where $\delta_{0,3}=-\delta_{1,2}=1$.

The tensor $Q_{\mu\nu}$ in the $n_2$--$n_3$ plane
also can be expanded in $\sigma$-matrices:
\begin{equation}
Q_{\mu\nu} = \sum_{k=0}^{3} Q_k \sigma_{k\mu\nu}\,,\quad
Q_k = \frac{1}{2} Q_{\mu\nu} \sigma_k^{\nu\mu}\,.
\label{Qdef}
\end{equation}
Using the Dirac equation, we obtain
\begin{equation}
\begin{split}
&Q_0 = - (x+y)\,,\quad
Q_1 = - i (x+y) \gamma_5 \hat{K} / 2\,,\\
&Q_2 = - (x+y) \gamma_5\,,\quad
Q_3 = - (x+y) + (x-y) \hat{K} / 2\,.
\end{split}
\label{Q}
\end{equation}
The conjugate tensor
$\overline{Q}_{\mu\nu} = \sum_{k=0}^{3} \delta_k Q_k \sigma_{k\nu\mu}$,
because $\overline{Q}_k = \delta_k Q_k$,
$\sigma_{k\mu\nu}^* = \sigma_{k\nu\mu}$.

The $e^\pm$ density matrices are
$\rho=\frac{1}{2}(\hat{p}-m)(1-\gamma_5\hat{a})$,
$\rho'=\frac{1}{2}(\hat{p}'+m)(1-\gamma_5\hat{a}')$.
Let's introduce two bases
\begin{equation}
\begin{split}
&e_0 = \frac{p}{m}\,,\quad
e_1 = \frac{(x+y-2)p - 2p'}{muv}\,,\quad
e_2 = \frac{2 w n_1 - (x-y) n_2}{uv}\,,\quad
e_3 = n_3\,;\\
&e'_0 = \frac{p'}{m}\,,\quad
e'_1 = \frac{(x+y-2)p' - 2p}{muv}\,,\quad
e'_2 = \frac{2 w n_1 + (x-y) n_2}{uv}\,,\quad
e'_3 = n_3\,;
\end{split}
\label{bases}
\end{equation}
where $u=\sqrt{x+y-4}$.
Then $a = \sum_{i=1}^{3} \zeta_i e_i$,
where in the c.~m.\ frame $\zeta_1$ is the longitudinal polarization,
$\zeta_2$ is the transverse polarization in the reaction plane,
and $\zeta_3$ is the transverse polarization perpendicular to this plane.
Introducing formally $\zeta_0=1$, we have
$\rho = \frac{1}{2} \sum_{i=0}^{3} \zeta_i \rho_i$,
where $\rho_0=\hat{p}-m$, $\rho_i=-\rho_0\gamma_5\hat{e}_i$.
Similarly, $\rho' = \frac{1}{2} \sum_{i'=0}^{3} \zeta'_{i'} \rho'_{i'}$,
where $\rho'_0=\hat{p}'+m$, $\rho'_i=-\rho'_0\gamma_5\hat{e}'_i$.

Finally, the cross section is
\begin{equation}
\frac{d\sigma}{dt\,d\varphi} = \frac{\alpha^2}{4 s^2 x^2 y^2}
\sum_{ii'jj'} F^{ii'}_{jj'} \xi_j \xi'_{j'} \zeta_i \zeta'_{i'}\,,
\label{csect}
\end{equation}
where
\begin{equation}
F^{ii'}_{jj'} = \sum_{kk'} \delta_{j'} \delta_{k'} \varepsilon_{j'}\,
\frac{1}{2} \Tr \sigma_{j'} \sigma_k \sigma_j \sigma_{k'}\,
\frac{1}{4} \Tr \rho'_{i'} Q_k \rho_i Q_{k'}
\label{F}
\end{equation}
(here the factor $\varepsilon_{0,1,3}=-\varepsilon_2=1$ appears,
in contrast to the Compton case~\cite{G},
because the photon with the momentum $k'$ is now initial instead of final,
and its density matrix has indices in the opposite order).
The right-hand side of~(\ref{csect}) depends on $\varphi$
because the polarizations are defined relative the reaction plane.
The final particles' polarizations $\zeta_i$, $\zeta'_{i'}$
describe probabilities of their registration by the detector;
when they are absent, $d\sigma=\frac{1}{4}d\sigma_{\text{unpol}}$~\cite{BLP}.
The cross section summed over the final particles' polarizations is
\begin{equation}
\frac{d\sigma}{dt\,d\varphi} = \frac{\alpha^2}{s^2 x^2 y^2} F\,,\quad
F = \sum_{jj'} F^{00}_{jj'} \xi_j \xi'_{j'}\,.
\label{sum}
\end{equation}
Polarizations of the final particles themselves are
\begin{equation}
\zeta^{(f)}_i = \frac{1}{F} \sum_{jj'} F^{i0}_{jj'} \xi_j \xi'_{j'}\,,\quad
\zeta^{(f)\prime}_{i'} = \frac{1}{F} \sum_{jj'} F^{0i'}_{jj'} \xi_j \xi'_{j'}\,.
\label{polar}
\end{equation}
The four-vectors of the final particles' polarizations are evidently
$a^{(f)} = \sum_{i=1}^{3} \zeta^{(f)}_i e_i$,
$a^{(f)\prime} = \sum_{i'=1}^{3} \zeta^{(f)\prime}_{i'} e'_{i'}$.
The components $F^{ii'}_{jj'}$ with $i\ne0$ and $i'\ne0$
describe the correlations of the final particles' polarizations.

We have calculated $F^{ii'}_{jj'}$ using REDUCE.
All nonzero components $F^{ii'}_{jj'}$ are presented in the Appendix
(where the notation $\overline{F}^{ii'}_{jj'}(x,y)=F^{ii'}_{jj'}(y,x)$
is used).

Repeating the derivation for the inverse process $e^+e^-\to\gamma\gamma$,
we can express its cross section via the same $F^{ii'}_{jj'}$~(\ref{F}):
\begin{equation}
\frac{d\sigma}{dt\,d\varphi} = \frac{\alpha^2}{4 s (s-4m^2) x^2 y^2}
\sum_{ii'jj'} F^{ii'}_{jj'} \varepsilon_j \varepsilon_{j'} \alpha_i \alpha_{i'}
\xi_j \xi'_{j'} \zeta_i \zeta'_{i'}\,,
\label{acsect}
\end{equation}
where $\alpha_0=-\alpha_{1,2,3}=1$.
This means that for the inverse process one has to substitute
$\xi_2\to-\xi_2$, $\xi'_2\to-\xi'_2$, $a\to-a$, $a'\to-a'$.
These substitutions follow from the change of signs of all the momenta
for the inverse process.
The cross section summed over the final photon polarizations
and the final photon polarizations are
\begin{equation}
\begin{split}
&\frac{d\sigma}{dt\,d\varphi} = \frac{\alpha^2}{s (s-4m^2) x^2 y^2} F\,,\quad
F = \sum_{ii'} F^{ii'}_{00} \alpha_i \alpha_{i'} \zeta_i \zeta'_{i'}\,,\\
&\xi^{(f)}_j = \frac{\varepsilon_j}{F} \sum_{ii'} F^{ii'}_{j0}
\alpha_i \alpha_{i'} \zeta_i \zeta'_{i'}\,,\quad
\xi^{(f)\prime}_{j'} = \frac{\varepsilon_{j'}}{F} \sum_{ii'} F^{ii'}_{0j'}
\alpha_i \alpha_{i'} \zeta_i \zeta'_{i'}\,.
\end{split}
\label{annihil}
\end{equation}

Comparison of the obtained functions $F^{ii'}_{jj'}$
with the corresponding functions for the Compton effect~\cite{G}
is more complicated because of the different choice
of fermion polarization unit vectors $e^{(\prime)}_{1,2}$.
However, for $i$, $i'=0$, 3 they are related by
\begin{equation}
\tilde{x} = -x\,,\quad
\tilde{y} = y\,,\quad
\tilde{v} = iv\,,\quad
\tilde{w} = iw\,,\quad
F^{ii'}_{jj'} = -i \varepsilon_j \tilde{F}^{i'j'}_{ij}\,,
\label{crossing}
\end{equation}
where the quantities of the Compton effect are denoted by tilde,
and the sign factor $\varepsilon_j$ corresponds to the substitution
$\xi_2\to-\xi_2$.

Now we discuss some limiting cases.
The first one is the threshold behavior ($x,y\to2$).
In this limit, the $e^+e^-$ pair is produced
with the orbital angular momentum $l=0$.
Since the initial photons can't have the total angular momentum $J=1$,
the $e^+e^-$ pair is produced with the total spin $S=0$,
and hence with the negative parity.
For pseudoscalar initial and final states,
the matrix element has the factorized form
$M\propto \vec{k}\cdot(\vec{e}\times\vec{e}\,') u(p') \gamma_5 v(p)$,
and
$|M|^2\propto(1+\xi_1\xi'_1+\xi_2\xi'_2-\xi_3\xi'_3)(1-\vec{a}\cdot\vec{a}\,')$
(where the last factor is $1+\zeta_1\zeta'_1+\zeta_2\zeta'_2-\zeta_3\zeta'_3$,
because $\vec{e}\,'_{1,2}=-\vec{e}_{1,2}$).

In the ultrarelativistic limit ($x,y\to\infty$),
$\frac{d\sigma}{dt}=\mathcal{O}\left(\frac{1}{s^2}\right)$,
and hence $F^{ii'}_{jj'}=\mathcal{O}(x^4,y^4)$.
However, if $e^+$ and $e^-$ have equal helicities
($\zeta_1=\zeta'_1=\pm1$, all the other components are zero),
then their production is suppressed due to helicity conservation,
and $\sum_{i,i'} F^{ii'}_{jj'} \zeta_i \zeta'_{i'} = \mathcal{O}(x^3,y^3)$
for all $j$, $j'$.

If $e^+$ and $e^-$ have opposite helicities
($\zeta_1=-\zeta'_1=\pm1$, all the other components are zero),
then the projection of their total angular momentum
onto their direction of motion is $\pm1$.
Projection of the total angular momentum of the photons
onto the collision axis is $\pm1\pm1=0$, $\pm2$.
Therefore, the reaction is forbidden at any energy,
when both the initial and the final particles move along the same line
in the c.~m.\ frame.
This means that $\sum_{ii'} F^{ii'}_{jj'} \zeta_i \zeta'_{i'}$
vanish at the kinematical boundaries $w=0$ for all $j$, $j'$.
Similarly, if $e^+$ and $e^-$ have equal helicities
($\zeta_1=\zeta'_1=\pm1$, all the other components are zero),
and the photons have opposite helicities
($\xi_2=-\xi'_2=\pm1$, all the other components are zero),
then the reaction is forbidden at the kinematical boundaries,
and $\sum_{ii'jj'} F^{ii'}_{jj'} \xi_j \xi'_{j'} \zeta_i \zeta'_{i'}$
vanishes at $w=0$.

Our results satisfy all these properties.
They are also consistent with the results of six independent calculations,
when the photons are either unpolarized or $e=e'=\pm n_3$,
and $e^\pm$ are unpolarized, have $a=a'=\pm n_3$, or $a=-a'=\pm n_3$.

We are indebted to the International Science Foundation
for partial financial support of V.~N.~B.\ (grant RP6000)
and A.~G.~G.\ (grant RAK000).

\section*{Appendix}

\begin{align*}
&F^{00}_{00} = F^{33}_{33} = x^3 y + 4 x^2 y - 4 x^2 + x y^3 + 4 x y^2 - 8 x y - 4 y^2\\
&F^{00}_{03} = F^{00}_{30} = F^{11}_{03} = F^{11}_{30}
= - F^{22}_{03} = - F^{22}_{30} = F^{33}_{03} = F^{33}_{30} = 4 v^2 w^2\\
&F^{01}_{02} = \overline{F}^{01}_{20} = \overline{F}^{10}_{02} = F^{10}_{20}
= - F^{23}_{13} = - \overline{F}^{23}_{31} = - \overline{F}^{32}_{13} = - F^{32}_{31} =\\
&\qquad{} - \left( x^2 y - 2 x^2 - x y^2 - 2 x y + 4 x + 4 y \right) v^3/u\\
&F^{02}_{02} = - F^{02}_{20} = F^{13}_{13} = - F^{13}_{31}
= - F^{20}_{02} = F^{20}_{20} = - F^{31}_{13} = F^{31}_{31}
= 4 v^2 w^3/u\\
&F^{00}_{11} = - F^{33}_{22} = - 2 (xy - 2 x - 2 y) x y\\
&F^{03}_{12} = - \overline{F}^{03}_{21} = - \overline{F}^{12}_{03} = F^{12}_{30}
= F^{21}_{03} = - \overline{F}^{21}_{30} = - \overline{F}^{30}_{12} = F^{30}_{21}
= 2 v^3 w x\displaybreak\\
&F^{00}_{22} = - F^{33}_{11} = - (x^2 + y^2) (xy - 2 x - 2 y)\\
&F^{01}_{23} = F^{01}_{32} = F^{10}_{23} = F^{10}_{32}
= - F^{23}_{01} = - F^{23}_{10} = - F^{32}_{01} = - F^{32}_{10}
= 4 v^3 w^2/u\\
&F^{02}_{23} = - \overline{F}^{02}_{32} = F^{13}_{01} = - \overline{F}^{13}_{10}
= - \overline{F}^{20}_{23} = F^{20}_{32} = - \overline{F}^{31}_{01} = F^{31}_{10}
= - 2 (y-2) v^4 w/u\\
&F^{00}_{33} = F^{33}_{00} =
- 2 \left( x^2 y^2 - 2 x^2 y + 2 x^2 - 2 x y^2 + 4 x y + 2 y^2 \right)\\
&F^{11}_{00} = - F^{22}_{33} = - \bigl( x^4 y - 2 x^4 + x^3 y^2 - 8 x^3 y + 8 x^3
+ x^2 y^3 - 20 x^2 y^2 + 40 x^2 y - 16 x^2 + x y^4\\
&\qquad{} - 8 x y^3 + 40 x y^2 - 32 x y
- 2 y^4 + 8 y^3 - 16 y^2 \bigr) / u^2\\
&F^{12}_{00} = F^{12}_{11} = - F^{12}_{22} = F^{12}_{33}
= F^{21}_{00} = F^{21}_{11} = - F^{21}_{22} = F^{21}_{33}
= 4 (x-y) v w^3/u^2\\
&F^{11}_{11} = F^{22}_{22} = 2 \left( x^3 y^2 - 2 x^3 y + 2 x^3 + x^2 y^3 - 2 x^2 y
- 2 x y^3 - 2 x y^2 + 2 y^3 \right) / u^2\\
&F^{11}_{22} = F^{22}_{11} = \left( x^4 y + x^3 y^2 - 4 x^3 + x^2 y^3 - 8 x^2 y^2 + 4 x^2 y
+ x y^4 + 4 x y^2 - 4 y^3 \right) / u^2\\
&F^{11}_{33} = - F^{22}_{00} = 2 \left( x^3 y^2 - 2 x^3 y + x^2 y^3 - 8 x^2 y + 8 x^2
- 2 x y^3 - 8 x y^2 + 16 x y + 8 y^2 \right) /u^2
\end{align*}

\end{document}